\documentclass[aps,prb,showpacs,twocolumn,amssymb,floats,epsfig]{revtex4}

\usepackage{graphicx}
\usepackage{epsfig}
\usepackage{amsfonts}
\usepackage{amsmath}
\usepackage{amssymb}
\usepackage{color}

\newcommand\ba{\begin{eqnarray}}
\newcommand\ea{\end{eqnarray}}
\newcommand\be{\begin{equation}}
\newcommand\ee{\end{equation}}
\newcommand\bi{\bibitem}
\newcommand{\ct}{\cite}

\def\non{\nonumber}

\def\al{\alpha}
\def\om{\omega}

\def\ep{\epsilon}
\def\ga{\gamma}

\def\si{\sigma}

\begin{document}

\title{Maximum group velocity in a one-dimensional model
with a sinusoidally varying staggered potential}
\author{Tanay Nag$^1$, Diptiman Sen$^2$, and Amit Dutta$^1$}
\affiliation{\small{$^1$Department of Physics, Indian Institute of Technology, 
Kanpur 208 016, India \\
$^2$Centre for High Energy Physics, Indian Institute of Science, Bengaluru
560 012, India}}

\begin{abstract}
We use Floquet theory to study the maximum value of the stroboscopic 
group velocity in a one-dimensional tight-binding model subjected to an 
on-site staggered potential varying sinusoidally in time. The results 
obtained by numerically diagonalizing the Floquet operator are analyzed using 
a variety of analytical schemes. In the low frequency limit we use adiabatic 
theory, while in the high frequency limit the Magnus expansion of the Floquet 
Hamiltonian turns out to be appropriate. When the magnitude of the staggered 
potential is much greater or much less than the hopping, we use degenerate 
Floquet perturbation theory; we find that dynamical localization occurs in 
the former case when the maximum group velocity vanishes. Finally, starting 
from an ``engineered" initial state where the particles (taken to be hard 
core bosons) are localized in one part of the chain, we demonstrate that the 
existence of a maximum stroboscopic group velocity manifests in a light cone 
like spreading of the particles in real space.
\end{abstract}


\maketitle

\section{Introduction}

In the real world, the speed at which information can propagate is limited 
by the speed of light; this results in the \textit{light cone} effect as 
postulated by the special theory of relativity. Is there a similar upper bound 
of the speed at which correlations (information) can propagate in interacting
quantum many-body systems? Following the seminal work
by Lieb and Robinson \ct{lieb72}, which established the existence of a 
maximum group velocity in a one-dimensional spin chain with a finite range 
interaction, some recent studies have explored this conjecture in several 
interacting many-body systems; these studies do indeed exhibit an effective 
light cone that sets a bound on the speed of propagation of correlations. 
This is reflected for example, in the growth of block entanglement entropy 
following a quench \ct{calabrese05}, or the collapse and revival of the 
Loschmidt echo \ct{stephan11, happola12}. The light cone like propagation of 
quantum correlations has also been observed experimentally by quenching a 
one-dimensional quantum gas in an optical lattice \ct{cheneau12}.

In parallel, there have been a plethora of studies of closed quantum systems 
driven periodically in time in the context of defect productions 
\ct{mukherjee08,mukherjee09}, dynamical freezing \ct{das10}, dynamical 
saturation \ct{russomanno12} and localization \ct{alessio13,bukov14,nag14}, 
dynamical fidelity \ct{sharma14}, and thermalization \ct{lazarides14} 
(for a review see Ref. \onlinecite{dutta15}). The study of periodically 
perturbed many-body 
systems has also gained importance because of the proposal of Floquet 
(irradiated) graphene \ct{gu11,kitagawa11,morell12}, Floquet topological 
insulators and the generation of topologically protected edge states 
\ct{kitagawa10,lindner11,jiang11,trif12,gomez12,dora12,cayssol13,liu13,tong13,
rudner13,katan13,lindner13,kundu13,basti13,schmidt13,reynoso13,wu13,manisha13,
perez1,perez2,perez3,reichl14,manisha14} some of which have been 
experimentally studied \ct{kitagawa12,rechtsman13,puentes14}. 

In this work, we use Floquet theory to explore the stroboscopic (i.e., 
measured at the end of each complete period) group velocity of a system of 
hard core bosons residing on a one-dimensional lattice in the presence of a 
staggered potential which is varying sinusoidally in time 
\ct{shirley65,griffoni98}. In particular we study the maximum value of the 
group velocity to observe the consequent light cone effect. Although the 
time-independent version of the model is integrable, the periodic sinusoidal 
perturbation renders the situation rather complicated since the corresponding 
Floquet operator cannot be obtained in a closed analytical form unlike the 
case of periodic perturbations which are piece-wise continuous in time 
\ct{nag14,dasgupta14}. One therefore has to use various approximation schemes 
valid in the appropriate regions of the parameter space to analyze the 
behavior of the stroboscopic group velocity.

\begin{figure*}[ht]
\begin{center}
\includegraphics[height=7.2cm,width=17.9cm]{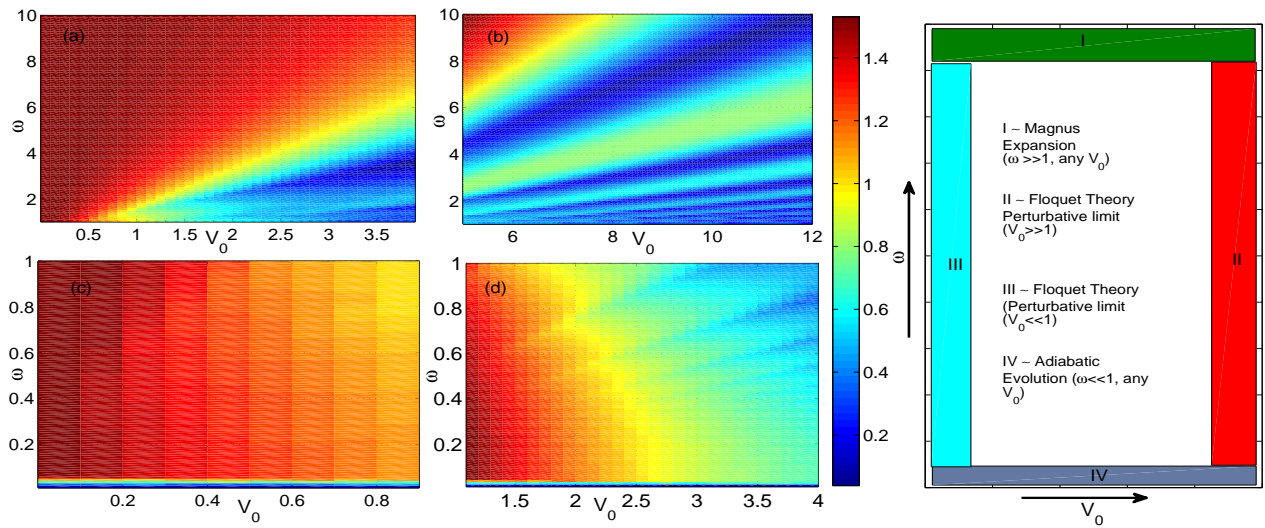}
\end{center}
\caption{(Color online) 
Density plots showing the variation of the maximum value of the stroboscopic 
group velocity as a function of the driving frequency $\om$ and the magnitude 
of the sinusoidally varying staggered potential $V_0$ (both in units of $\ga 
=1$). In (a) $V_{\rm max}$ is always finite (red 
region) for large $\om$ and small $V_0$ (top left corner). But there are some 
regions where $V_{\rm max}$ becomes very small (blue region) in the 
intermediate $\om$ range and for large $V_0$ (bottom right corner). (b) 
shows that there exists an array of regions where $V_{\rm max}$ vanishes 
(dark blue regions). The line-like regions of zero velocity are not equispaced 
in $\om$ for a given value of $V_0$. (c) shows that $V_{\rm max}$ never 
becomes zero for small $\om$ and small $V_0$, although $V_{\rm max}$ is very 
small for extremely low frequencies (blue region near the horizontal axis). 
(d) shows that $V_{\rm max}$ decreases as one increases $V_0$ while keeping 
$\om$ fixed at small values. At extremely small frequencies, $V_{\rm max}$ 
shows behavior similar to that in (c). (e) A 
schematic diagram showing the validity of the various limits of the Floquet 
theory in different regions in the $\om-V_0$ plane.} \label{lce} \end{figure*} 

The paper is organized in the following way. In Sec. \ref{model}, we present 
the Hamiltonian of the model under consideration and discuss the generic 
behavior of the maximum value of the stroboscopic group velocity 
$V_{\rm max}$ as a function of the amplitude and frequency of the periodic 
perturbation; this is derived using the numerically obtained Floquet 
quasi-energies. In Sec. \ref{adiabatic}, we use adiabatic theory to find the 
behavior of $V_{\rm max}$ in the low frequency limit while the high frequency 
limit is treated within a Magnus expansion in Sec. \ref{magnus}. In Sec. 
\ref{large_V} (Sec. \ref{small_V}), we use a Floquet perturbation 
theory \ct{soori10} when the hopping term is much smaller (greater) than the 
amplitude of the staggered potential. In the case of a large magnitude of the 
staggered potential, we point to the situations when the maximum stroboscopic 
group velocity vanishes resulting in the so-called \textit{dynamical 
localization}. We demonstrate a light cone like propagation of particles in 
real space in Sec. \ref{light_cone}. Concluding remarks are presented in 
Sec. \ref{concluding}.

\section{Model and the stroboscopic group velocity}
\label{model}

We consider a Hamiltonian of the tight-binding form
\be H = -\ga \sum_{l=1}^L (b_l^{\dagger}b_{l+1} + b_{l+1}^\dagger b_l), 
\label{eq:ham} \ee
where $\ga$ is the hopping amplitude, and
$b_l$ denotes bosonic annihilation operators defined on a one-dimensional
lattice satisfying the hard core condition $(b_l^{\dagger})^2 = (b_l)^2 =0$.
The single particle dispersion is given by $E_k = - 2 \ga \cos k$. This model 
describes a system of hard core bosons in a gapless superfluid phase 
\ct{klich07}. (By the Jordan-Wigner transformation \ct{lieb61}, this 
system is equivalent to one with
spinless fermions). When the system is perturbed by a spatially alternating 
potential $V_l = (-1)^l V_0$, a gap opens up in the spectrum for any non-zero 
value of $V_0$ thereby driving it to a gapped Mott insulator phase. Our aim 
here is to investigate the response of the system subjected to an alternating
potential varying sinusoidally in time as $V(t)=V_0\sin(\om t)$. We will
analyze the behavior of the stroboscopic group velocity, measured after $n$ 
complete periods of the driving, and its maximum value $V_{\rm max}$ as a 
function of the driving frequency $\om$ and the amplitude $V_0$.

In the presence of the alternating potential, the Hamiltonian in 
Eq.~\eqref{eq:ham} reduces in momentum space to a $2 \times 2$ matrix form
in terms of the momenta $k$ and $k+\pi$,
\be H_k(t)=-2 \ga \cos k ~\si^z +V_0\sin(\om t) ~\si^x, \label{eq:ham_k} \ee
where $\si^{x,z}$ denote pseudo-spin Pauli matrices. Clearly the spectrum is 
gapless for $V_0 =0$. We will set $\ga =1$, Planck's 
constant $\hbar =1$ and the lattice spacing equal to 1 in the rest of the 
paper. Hence $\om$, $V_0$, $k$ and $V_{\rm max}$ (to be defined below) will 
all be dimensionless.

Using the Jordan-Wigner transformation, the time-independent 
part of the Hamiltonian can be mapped to a system of spinless fermions on a 
one-dimensional lattice with a hopping amplitude $\ga$, while the 
time-dependent part of the Hamiltonian corresponds to a staggered 
chemical potential which is sinusoidally driven with a frequency $\om$. 
This equivalence between the hard core bosonic and the 
spinless fermionic models does not hold in higher than one dimension. 
Nevertheless, in higher dimensions there are non-interacting fermionic models 
with a sinusoidally driven chemical potential for which the behavior of the 
stroboscopic group velocity and the dynamical localization is similar to 
what is reported here.

Defining the time period $T=2\pi/\om$, the stroboscopic Floquet operator for 
each momentum mode is given by the unitary operator
${\cal F}_k (V_0, T) = {\cal T}\exp \left(-i \int_0^T dt H_k(t) \right)$, 
where ${\cal T}$ denotes the time-ordering. This operator 
cannot be computed analytically for a sinusoidal driving. One can however 
numerically calculate ${\cal F}_k$ and find its eigenvalues which take the form
$\exp(-i \mu_k^{\pm} T)$ where $\mu_k^{\pm} =\pm \mu_k$ are the quasi-energies.
The group velocity can be obtained from the quasi-energies as 
\be v_k = \partial \mu_k/\partial k. \label{eq_group_velocity} \ee
The maximum of $v_k$ as a function of $k$ gives the quantity $V_{\rm max}$, 
which is the main object of interest in this paper. The 
physical interpretation of the stroboscopic group velocity is that if the 
quantum correlations are measured only at the end of each complete period, 
they would appear to propagate with a maximum speed $V_{\rm max}$. This 
quantity is presented in Fig.~\ref{lce} as a function of $V_0$ and $\om$. 

Upon inspecting the results presented in Fig.~\ref{lce}, one finds that 
$V_{\rm max}$ tends to saturate at some value for large values of $\om$ and 
small or intermediate values of $V_0$ ($V_0 \le 1$) [see Fig.~\ref{lce} (a)]. 
The maximum group velocity shows an interesting behavior when both $V_0$ and 
$\om$ become large, as shown in Fig.~\ref{lce} (b). In this limit, one finds 
that for a given frequency $\om$, $V_{\rm max}$ vanishes in regular intervals
of $V_0$. On the other hand, for a given $V_0$, the zeros of $V_{\rm max}$ lie 
in increasing intervals of $\om$. In this regime $V_{\rm max}$ is given by 
the zeros of a Bessel function as we will show below. But when $\om$ is small, 
$V_{\rm max}$ never becomes zero, although in the limit $\om \to 0$, 
$V_{\rm max}$ becomes very small irrespective of $V_0$; see the bottom regions
of Figs.~\ref{lce} (c) and (d). The maximum group velocity gradually
decreases with $V_0$ if one keeps $\om$ fixed at a lower value, as shown in
Fig.~\ref{lce} (d). In subsequent sections, we will use different
analytical methods to analyze the various behaviors described here.

Finally, we note that we will consider the entire Brillouin 
zone ranging from $k=-\pi/2$ to $k=\pi/2$. The time-dependent part of the 
Hamiltonian in Eq.~\eqref{eq:ham} is quantum critical and gapless in the 
thermodynamic limit for the modes with $k=\pm \pi/2$. The 
minimum frequency scale of the bare tight-binding Hamiltonian (with $\hbar=1$)
is determined by the system size ($\sim 1/L$), while the maximum frequency 
scale appears for the modes $k=0, ~\pi$ and is equal to $2$ since $E_k = 
- 2 \cos k$.

\section{Adiabatic limit of low frequency}
\label{adiabatic}

The behavior of the quasi-energy can be explained in the low frequency 
limit where the adiabatic theory holds. We will choose the basis states as 
the eigenstates of the pseudo-spin operator $\si^z$, i.e., $(1~0)^T$ and 
$(0~1)^T$. In this limit, the product of the time period $T$ and the Floquet 
quasi-energies $\mu_k^{\pm}$ is equal to the dynamical phase $\ep_k^{\pm}$ 
accumulated over a complete time period $T$; this is given by
\begin{widetext}
\be \ep_k^{\pm}= \pm \int_0^{2\pi/\om}~dt~\sqrt{4\cos^2 k+V_0^2 \sin^2 \om t} 
~=~ \pm \left[\frac{2\sqrt{4 \cos ^2 k+V_0^2}}{\om}~ E\left(\frac{V_0^2}{V_0^2
+4 \cos ^2 k}\right)+\frac{4\cos k}{\om}~ E\left(-\frac{1}{4} V_0^2 \sec ^2 k
\right)\right], \label{dp} \ee
\end{widetext}
where $E(x)$ is the elliptic integral \ct{gradshteyn}. [We note that the Berry 
phase term vanishes in this problem since the closed path traced out by the 
Hamiltonian in \eqref{eq:ham_k} as $t$ goes from zero to $T$ is a line in the 
$x-z$ pseudo-spin space; such a line covers zero solid angle at the origin 
$(x,y,z)= (0,0,0)$.] Interestingly, the behavior of the quasi-energy can be 
qualitatively explained up to certain values of $V_0$. Let us elaborate 
on this below.

\begin{figure}[ht]
\begin{center}
\includegraphics[height=6.0cm]{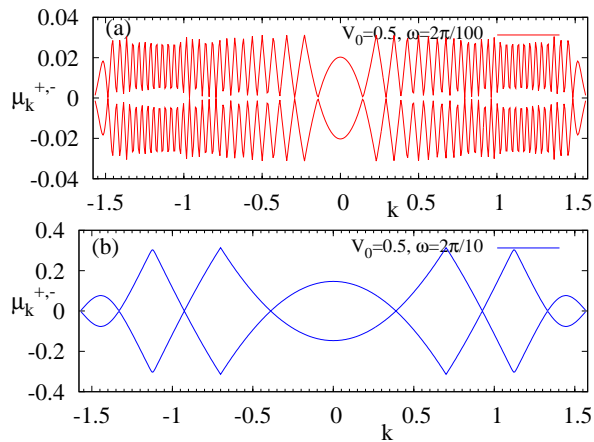}
\end{center}
\caption{(Color online) 
Plots showing the variation of the quasi-energy $\mu_k^{\pm}$ as a 
function of $k$ with the potential for $V_0<1$. In this parameter regime we 
choose two cases: (a) $V_0>\om$ and (b) $V_0<\om$. The behavior observed 
in (a) can be explained using adiabatic theory with small $V_0$. The 
locations of the quasi-degenerate points in (b) are discussed in Sec. 
\ref{small_V}.} \label{qe_lv} \end{figure}

In the limit $V_0 \ll 1$ and very small $\om$, we can use 
the form of the elliptic functions to reduce Eq.~\eqref{dp} to
\be \ep_k^{\pm} \simeq \pm \left[ \frac{4 \pi \cos k}{\om} + \frac{\pi V_0^2 
\sec k}{4 \om} + O(V_0^2)\right]. \label{dplv} \ee
A quasi-degeneracy occurs when this dynamical phase $\ep_k^{\pm}=m \pi$. This 
condition successfully gives the number of quasi-degenerate points along with 
values of the quasi-degenerate momenta, $k=\pm ~{\rm arccos}[(m\om +
\sqrt{-4 V_0^2+m^2\om^2})/8]$ [see Fig.~\ref{qe_lv} (a)]. (We 
note that the critical modes $k =\pm \pi/2$ are always quasi-degenerate.)

In the intermediate potential range $V_0\sim 1$, the behavior of the 
quasi-energy is again determined by the adiabatic evolution of two-level 
systems. The number of quasi-degenerate points is successfully given by 
$\ep_k^{\pm}=m\pi$, where $\ep_k^{\pm}$ is given by
\be \ep_k^{\pm} \simeq \pm \left[\frac{4 \cos ^2 k}{V_0 \om}+\frac{8 
\cos^2 k \log(4 V_0)}{V_0 \om}+\frac{4 V_0}{\om} \right]. \label{dphv} \ee

\begin{figure}[ht]
\begin{center}
\includegraphics[height=6.0cm]{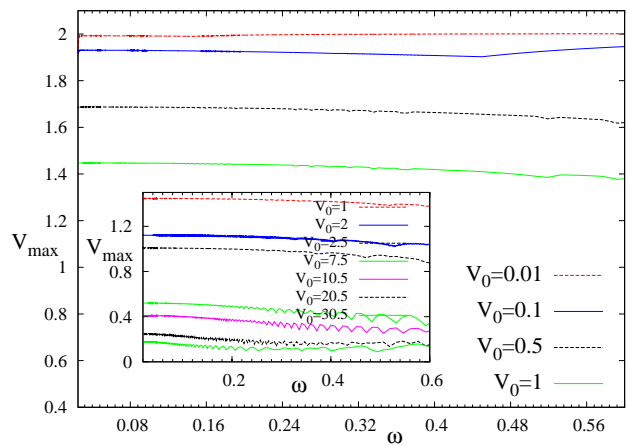}
\end{center}
\caption{(Color online) 
Plot showing that the maximum group velocity $V_{\rm max}$ is nearly 
independent of the frequency (or slightly decreasing with increasing $\om$) 
at low frequency for both small and large values of $V_0$. $V_{\rm max}$ 
decreases as $V_0$ increases. The inset shows that $V_{\rm max}$ exhibits some 
wiggles with $\om$ for high values of $V_0$.} \label{gv_w_1} \end{figure}

The group velocity can be obtained from Eqs.~\eqref{dplv} and
\eqref{dphv}; in the former case, we find $v_k\simeq \pm (2\sin k - (V_0^2 
\sec k \tan k)/8 + O(V_0^2))$ while in the latter $v_k\simeq \pm(2\sin (2k)/
(V_0 \pi)+4 \sin (2k) \log(4V_0)/(V_0 \pi))$. Therefore, we conclude that 
$V_{\rm max}$ is (nearly) independent of $\om$ for small frequencies; see 
Fig.~\ref{gv_w_1}. 
We find that $V_{\rm max}= 2$ for very small $V_0$ while for $V_0 \gg 1$, 
$V_{\rm max} \sim 1/ V_0$ as shown in Fig.~\ref{gv_v} for small values of 
$\om$. In the limit of small $V_0$ and $\om$ ($\om < 
1/L$), the system hardly senses the periodic driving and hence $V_{\rm max}$ 
is determined by the bare tight-binding Hamiltonian; in the limit of large
$V_0$, the Floquet perturbation theory holds (see Sec. \ref{large_V}) which
explains the $1/V_0$ decay of $V_{\rm max}$.

\section{Magnus expansion for high frequency}
\label{magnus}

The time evolution operator describing the Schr\"odinger evolution of a 
quantum system is given by ${\cal T} \exp[-i\int_0^t dt' H(t')] = \exp
[\Lambda(t)]$. In the Magnus expansion, the operator $\Lambda(t)$ is 
decomposed in the form $\Lambda(t) = \sum_{k=1}^{\infty} \Lambda_k(t)$ 
\ct{blanes09}. The advantage of using a Hamiltonian periodic in time is that 
the stroboscopic unitary operator (i.e., the Floquet operator) can be 
expressed in the form $F(T) = \exp(-i H_F T)$, where $H_F$ is the 
corresponding Floquet Hamiltonian. Thus, for a periodically driven system, 
the Magnus expansion enables us to express the Floquet Hamiltonian in the form 
$H_F = \sum_{n=1}^{\infty} H_F^{(n)}$, where the $H_F^{(n)}$'s
can be expressed in the following form \ct{alessio13}:
\ba H_F^{(1)}&=&\frac{1}{T}\int_0^T~dt H(t), \non\\
H_F^{(2)}&=& -~ \frac{i}{2T}\int_0^T~dt_1~\int_0^{t_2}~dt_2 \left[H(t_1),H(t_2)
\right], \non\\
H_F^{(3)}&=& -~ \frac{1}{6T}\int_0^T~dt_1~\int_0^{t_2}~dt_2 \int_0^{t_2}~dt_3 
\non\\
&& ~~~~~~~~\left(\left[H(t_1),\left[H(t_2),H(t_3)\right]\right]+
1\leftrightarrow 3 \right). \label{meh} \ea 
As shown below, the $n$th-order term decreases as $1/\om^{n-1}$, and is 
therefore vanishingly small for $\om \to \infty$.
For the model given by Eq.~\eqref{eq:ham_k}, $H_{F,k}^{(1)}=\al \si^z$, 
$H_{F,k}^{(2)}=2V_0\al/\om~\si^y$ and
$H_{F,k}^{(3)}=-(4\al^2 V_0\pi/3\om^2)~\si^x -(3\al V_0^2/\om^2)~\si^z$.

We will work in the high frequency limit and retain terms up to the order 
$1/\om$. We then arrive at an effective Hamiltonian given by 
\ba H_{F,k}^{\rm eff} = (\alpha-\frac{3 \alpha V_0^2}{\om^2})~\si^z
+\frac{2 \alpha V_0}{\om}~\si^y -\frac{4\alpha^2 V_0 \pi}{3 \om^2}~\si^x ,
\label{mehf} \ea
where $\alpha=-2\cos k$. The effective quasi-energies, 
obtained by diagonalizing \eqref{mehf} and retaining terms of order 
$(V_0/\om)^2$, 
are found to be $\mu_{\rm eff}^{\pm}= \pm \sqrt{ \alpha^2 -2 \alpha^2
V_0^2/ \om^2} \simeq \pm \alpha (1 -V_0^2/\om^2)$; the quasi-degeneracy points
occur at $k =\pm \pi/2$ where $\al$ vanishes. In the limit $\om \to \infty$, 
$\mu_{\rm eff}^{\pm}\approx \pm \alpha$ . Therefore, the maximum group
velocity $V_{\rm max}$ becomes $2$ irrespective of $V_0$. This can also be 
explained simply by noting that the periodically varying perturbation in 
the Hamiltonian \eqref{eq:ham_k} vanishes on average in the high frequency 
limit. Moreover, for smaller values of $V_0$, $V_{\rm max}$ reaches its 
saturation value $V_{\rm max}=2$ at a smaller value of $\om$ as compared to 
higher values of $V_0$. 

\section{Large potential compared to the hopping amplitude: $V_0 \gg \ga$}
\label{large_V}

Let us now examine the case where the hopping amplitude $\ga$ can be treated
as a perturbing parameter in the Hamiltonian given by Eq.~\eqref{eq:ham_k}.
It is useful to consider a unitary transformation which shifts the time 
dependence of the Hamiltonian to the diagonal term, so that the transformed 
Hamiltonian takes the form $H_k(t)=V_0 \sin(\om t)\si^z+2\ga \cos k \si^x$.
(We now set $\ga = 1$ as usual). The time-dependent Schr\"odinger equation in 
the new basis can then be written as $i|\dot\phi_k^{\pm}(t)\rangle=H_k(t)|
\phi_k^{\pm}(t)\rangle$. Dividing both sides of the equation by $V_0$ and 
rescaling $t$ to $tV_0$, the Schr\"odinger equation can be rewritten as
\ba i|\dot\phi^+_k(t)\rangle &=& \sin (\frac {\om}{V_0} t) |\phi^+_k(t)
\rangle + \frac {2 \cos k}{V_0} |\phi^-_k(t) \rangle, \non \\
i|\dot\phi^-_k(t)\rangle &=& - \sin (\frac {\om}{V_0} t) |\phi^-_k(t)
\rangle + \frac {2 \cos k}{V_0} |\phi^+_k(t) \rangle. \ea
We will set $\om/V_0=a $ and $2 \cos k/V_0=b$ in subsequent calculations.
The solutions in the zeroth order of $w$ are given by 
\ba |\phi_k^+(t)\rangle&=&\left( \begin{array}{cc} 
c_k^+(0)~e^{i\cos (at)/a} \\ 
0 \end{array} \right), \non\\
|\phi_k^-(t)\rangle&=&\left( \begin{array}{cc} 
0 \\ 
c_k^-(0)~e^{-i\cos(at)/a} \end{array} \right), \label{fphv1} \ea
where $c_k^{\pm}(0)$ denote the probability amplitudes of the states 
$|\phi_k^{\pm}\rangle$ at time $t=0$. We find that $|\phi_k^+(T)\rangle
=|\phi_k^+(0)\rangle$ and $|\phi_k^-(T)\rangle=|\phi_k^-(0)\rangle$, 
implying that these solutions are degenerate in Floquet theory. 
We therefore employ a degenerate perturbation theory to include the hopping 
term perturbatively and find the time-dependent coefficients $c_k^{\pm}(t)$, 
which satisfy the evolution equations
\ba i\dot c_k^+(t) &=& bc_k^-(t)~e^{-i2\cos(at)/a},\non\\
i\dot c_k^-(t) &=&bc_k^+(t)~e^{i2\cos(at)/a}. \label{fphv2} \ea
To incorporate the correction up to first order in the hopping, we substitute 
$c_k^{\pm}(t)$ appearing on the right sides of Eqs.~\eqref{fphv2} by 
$c_k^{\pm}(0)$, respectively. The solution at time $t=T=2\pi/a$ is then given 
by
\ba c_k^+(T) &=& c_k^+(0)-(i2 \pi bc_k^-(0)/a)~J_0(2V_0/\om), \non\\
c_k^-(T) &=& c_k^-(0)-(i2 \pi bc_k^+(0)/a)~J_0(2V_0/\om). \label{fphv3} \ea

Up to first order in the hopping $\ga$, the Floquet operator is given by 
\be F_k(T)=\left[ \begin{array}{cc} 
1 & -(i2 \pi b/a) J_0(2V_0/\om) \\
-(i2 \pi b/a) J_0(2V_0/\om) & 1 \end{array} \right], \label{fphv4} \ee
so that $(c_k^+(T)~c_k^-(T))^T = F_k(T) (c_k^+(0)~c_k^-(0))^T = \exp(i 
\theta_k^{\pm})(c_k^+(0)~c_k^-(0))^T$, where $(... ~...)^T$ denotes transpose.
Diagonalizing matrix \eqref{fphv4}, one obtains the eigenvalues
$e^{i\theta_k^{\pm}}$, and hence $\theta_k^{\pm}=\mu_k^{\pm}T=
\pm (2 \pi b /a)~J_0(2V_0/\om) = 2 \cos k J_0(2V_0/\om)$.

\begin{figure}[ht]
\begin{center}
\includegraphics[height=5.6cm]{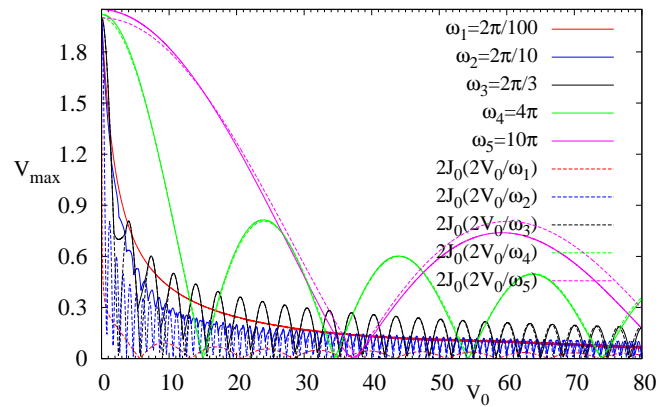}
\end{center}
\caption{(Color online)
Variation of the maximum group velocity as a function 
of $V_0$. For higher values of $V_0$ and frequency $\om$, the numerically 
obtained $V_{\rm max}$ is found to match the Bessel function given by 
$2 ~J_0(2V_0/\om)$. On the other hand, for very small values of the frequency 
$V_{\rm max}$ does not match the Bessel function.} \label{gv_v} \end{figure}

For large $V_0$ and large $\om$, the quasi-energy is given by
$\mu_k^{\pm} \simeq \pm 2 \cos k \sqrt{\om/\pi V_0}\cos (2V_0/\om-\pi/4)$.
The maximum group velocity is $V_{\max}=2 \sqrt{\om/\pi V_0}\cos (2V_0/\om
-\pi/4)$ which that vanishes at $2 V_0/\om=(n+3/4)\pi$. This matches 
the observed numerical results presented in Fig.~\ref{gv_v}. We note that 
the maximum group velocity vanishes when $J_0 (2V_0/\om)=0$. 

Furthermore, given $\mu_k^{\pm} =\pm 2\cos k J_0(2V_0/\om)$, one can find the 
values of the momenta for which quasi-degeneracies occur in the Floquet 
spectrum given by $\mu_k T=m\pi$, namely, $k={\rm arccos}[m\om/(4J_0(2V_0/
\om))]$. A solution for $k=\pm\pi/2$ can be found only for $m=0$. This 
behavior is identical to that obtained from the Magnus expansion for $\om \gg 
V_0$. From Fig.~\ref{gv_v}, we find that for high $V_0$, Floquet 
perturbation theory works better at higher frequencies where $V_{\rm max}
=2 J_0(2V_0/\om)$. We note that for very small values of $\om$, $V_{\rm max}$ 
falls off as $1/V_0$ as predicted by the adiabatic theory. Therefore, a 
crossover in the behavior of $V_{\rm max}$ as a function of $V_0$ and $\om$ 
is expected. The crossover happens between two types of behaviors of the
maximum group velocity, i.e., $V_{\rm max}\propto V_0^{-1}$ and $V_{\rm max}
\propto 2J_0(2V_0/\om)$. Although $V_{\rm max}$ never becomes zero (but shows 
a dip) at the zeros of a Bessel function for small $\om$, we find that 
$V_{\rm max}$ indeed vanishes at these points for higher values of $\om$.

The vanishing of $V_{\rm max}$ when $J_0(2V_0/\om_0)=0$, 
corresponds to the coherent destruction of tunneling 
\ct{grossmann91,kayanuma94} or dynamical freezing \ct{das10}. When the bare 
energies (diagonal terms) of a two-level system are sinusoidally driven with 
a driving frequency which is much larger than the tunneling (appearing 
in the off-diagonal terms of the corresponding $2 \times 2$ Hamiltonian), 
the system may get frozen in its initial state even though the dynamics is 
perfectly unitary; this is known as the coherent destruction of tunneling 
which occurs when the transition 
probability to the other state given by $J_0(2V_0/\om_0)$ vanishes. The 
present model ideally represents such a situation when $V_0 \gg 1$ and 
$\om_0 \gg 1$ with $V_0/\om_0 \sim O(1)$ as also observed numerically 
in Fig.~\ref{gv_v}. Whenever $V_{\rm max}$ vanishes, the quantum 
correlations do not propagate as we will discuss in Sec. \ref{light_cone};
this also leads to a real space localization of hard core bosons.

\section{Small potential compared to the hopping amplitude: $V_0 \ll \ga$}
\label{small_V}

We now consider the other limit, $V_0 \ll \ga = 1$, when $V_0$ can be treated 
perturbatively. At the zeroth order in $V_0$, the Hamiltonian \eqref{eq:ham_k} 
reduces to $H_k=-2\cos k~\si^z$ with eigenfunctions
\be |\phi_k^+(t)\rangle=\left( \begin{array}{cc} 
e^{-i2\cos k~t} \\ 0 \end{array} \right), ~~~
|\phi_k^-(t)\rangle=\left( \begin{array}{cc} 
0 \\ ~e^{i2\cos k~t} \end{array} \right). \label{fplv1} \ee
In the non-degenerate case when $e^{-i2\cos k~T} \ne e^{i2\cos k~T}$, 
$T=2\pi/\om$, it can be easily shown that the first order correction in the 
quasi-energy vanishes since $\langle \si^x\rangle=0$ when the expectation
values are calculated with the eigenfunctions in Eq.~\eqref{fplv1}. 
This necessitates the application of a degenerate perturbation theory 
\ct{soori10} when the condition $e^{-i2\cos k~T} = \pm 1$ is satisfied, 
implying that $4 \cos k = m \om$. We will distinguish between two situations, 
$4 \cos k /\om \ne 1$ and $4 \cos k /\om = 1$; as we will show below, in the 
former case there is a correction of order $V_0^2$, while
in the latter case a correction of order $V_0$ emerges.

Let us first discuss the situation in which $4 \cos k/\om \ne 1$. In the 
same spirit as in Sec. \ref{large_V}, the quasi-states are chosen to be
\ba |\phi_k^+(t)\rangle &=& \left( \begin{array}{cc} 
c_k^+(t)~e^{-i2\cos k~t} \\ 0 \end{array} \right), \non\\
|\phi_k^-(t)\rangle &=& \left( \begin{array}{cc} 
0 \\ c_k^-(t)~e^{i2\cos k~t} \end{array} \right). \label{fplv2} \ea
We note that $|\phi_k^+(T)\rangle=e^{-i2T\cos k}|\phi_k^+(0)\rangle$ 
and $|\phi_k^-(T)\rangle=e^{i2T\cos k}|\phi_k^-(0)\rangle$.
The time-dependent coefficients 
satisfy the Schr\"odinger equation 
\ba i\dot c_k^+(t) &=&-iV_0\sin(\om t) ~c_k^-(t) ~e^{-i4t \cos k}, \non \\
i\dot c_k^-(t) &=&-iV_0\sin(\om t) ~c_k^+(t) ~e^{i4t \cos k}. \label{fplv3} \ea
Within the first order perturbative approximation, we substitute $c_k^{\pm}(t)
= c_k^{\pm}(0)$ on the right hand side of the above equations. At $t=T$, we 
find that
\ba c_k^+(T)&=&c_k^+(0)-\frac{\om V_0 c_k^-(0)}{\om^2-16\cos^2 k}\sin
(4T\cos k), \non\\
c_k^-(T)&=&c_k^-(0)+\frac{\om V_0 c_k^+(0)}{\om^2-16\cos^2 k}\sin
(4T\cos k). \label{fphl4} \ea
Considering first the situation $4\cos k\ne \om$,
the Floquet operator up to the first order in $V_0$ at time $t=T$ is given by 
\begin{widetext}
\be F_k(T)=\left[ \begin{array}{cc} 
e^{-i2T\cos k} & -\frac{\om V_0}{\om^2-16\cos^2 k} \sin (4T\cos k) 
e^{-i2T\cos k} \\
\frac{\om V_0}{\om^2-16\cos^2 k} \sin (4T\cos k) e^{i2T\cos k} & 
e^{i2T\cos k} \end{array} \right]. \label{fplv5} \ee
\end{widetext}
Diagonalizing the Floquet operator in Eq.~\eqref{fplv5}, we get the Floquet 
quasi-energies $\exp(i\mu_k^{\pm} T)=\cos(2T\cos k) \pm i\sqrt{\sin^2 
(2T\cos k) +[\om V_0\sin (4T\cos k)/(\om^2-16\cos^2 k)]^2}$, and hence 
$\mu_k^{\pm} =2\cos k+O(V_0^2)$. The maximum group velocity becomes equal 
to $2$ for small $V_0$. 

On the other hand, when $4\cos k = \om$, we have $|\phi_k^+(T) \rangle=
|\phi_k^+(0)\rangle$ and $|\phi_k^-(T)\rangle=|\phi_k^-(0)\rangle$ to zeroth 
order in $V_0$. Solving the Schr\"odinger equations within the first order 
approximation, one finds that the time-dependent coefficients are given by
\be c_k^+(T)=c_k^+(0)+\frac{\pi V_0 c_k^-(0)}{\om},~~
c_k^-(T)=c_k^-(0)-\frac{\pi V_0 c_k^+(0)}{\om}. \label{fplv6} \ee
The Floquet operator is given by 
\be F_k(T)=\left[ \begin{array}{cc} 
1 & V_0 \pi/\om \\
-V_0 \pi/\om & 1 \end{array} \right]. \label{fplv7} \ee
The eigenvalues of the Floquet operator are $e^{i\mu_k^{\pm} T}=
1\pm iV_0\pi/\om$. The quasi-energy $\mu_k^{\pm}={\log} 
(1\pm iV_0\pi/\om)/T \approx \pm V_0/2$, leading to a first-order correction 
to the quasi-energy unlike the previous case $4 \cos k \ne \om$. We also find 
that the quasi-degenerate momentum modes are given by $k=\pm ~{\rm arccos}(m 
\om/4)$. Referring to Fig.~\ref{qe_lv} (b) for $V_0=0.5$ and $\om=2\pi/10$, we
note that the number of quasi-degenerate points is successfully predicted by 
this theory.
A correction to the quasi-energy of the order of $V_0$ appears at only $m=1$. 

\section {Light cone like propagation of particles in real space with 
stroboscopic time $T$}
\label{light_cone}

\begin{figure*}[ht]
\begin{center}
\includegraphics[height=8.3cm,width=17.2cm]{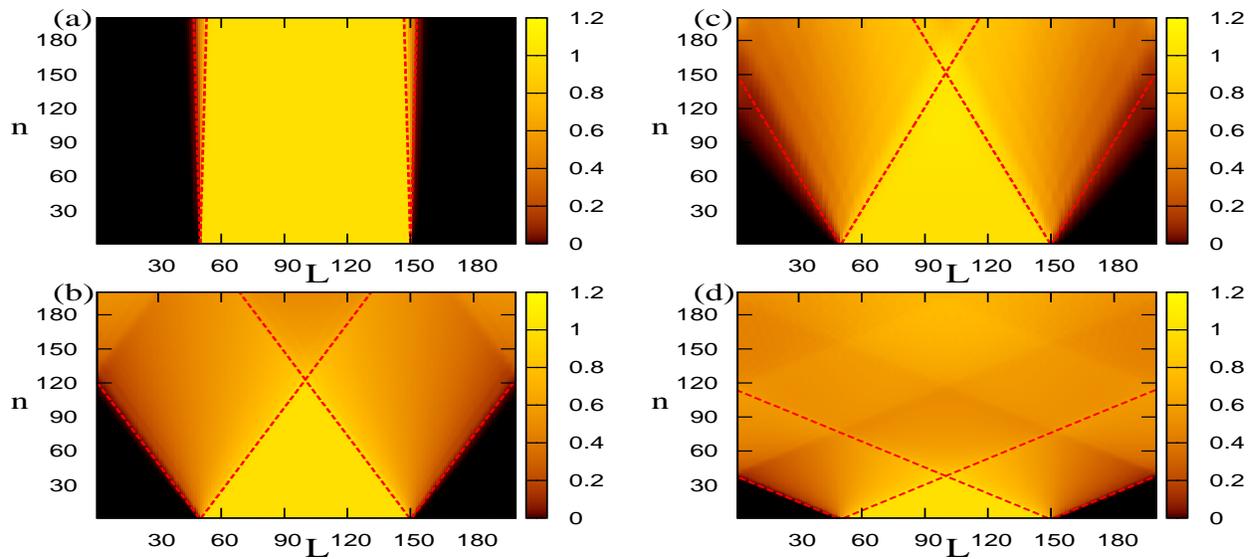}
\end{center}
\caption{(Color online)
Density of particles as a function of 
the stroboscopic time $nT$ and lattice site $L$. The red dotted line signifies 
the localization-delocalization boundary. One can determine the maximum group 
velocity from the slope of this line [$V_{\rm max}=dL/( Tdn)$], which is found 
to agree well with $V_{\rm max}$ obtained analytically from a Bessel 
function. Fig. (a) depicts a situation where a dynamical localization 
(dynamical freezing) nearly happens, for $\om=4\pi$ and $V_0=15$; 
$V_{\rm max}$ nearly vanishes when $J_0(2V_0/\om)=0$. 
(b) shows that particles move with a maximum group
velocity of $V_{\rm max}=0.814$ for $\om=4 \pi$ and $V_0=24$. (c) and 
(d) show that no dynamical localization is observed for $\om=2\pi/3$ with 
$V_0=12$ and $V_0=14$, respectively.} \label{lce_1} \end{figure*} 

In the earlier sections, we discussed the maximum group velocity 
$V_{\rm max}$ for a given set of parameter values $V_0$ and $\om$ as 
presented in Fig.~\ref{lce}. Here we illustrate how the light cone effect 
arising due to the existence of an upper bound to the group velocity 
manifests in the real space propagation of particles as shown in 
Fig.~\ref{lce_1}; we see that there is a dynamical localization when 
$V_{\rm max} \to 0$. This is illustrated by choosing an initial state at $t=0$
of a $200$-site system in which the sites labeled $51$ to $150$ are filled 
(shown by the light region) and the remaining sites are empty (shown by the 
dark region); this initial state evolves with the total Hamiltonian, i.e.,
the tight-binding part as well as the sinusoidal driving of the staggered 
potential. For every stroboscopic instant ($t=nT$), we can find the particle 
density at each site by numerically studying the time evolution of the 
initial density matrix $\rho(0)$, namely, $\rho(nT)=F(nT)\rho(0)F^{\dagger}
(nT)$, where $F(nT)$ is the real space Floquet operator at time $t=nT$. 
The slope of the red dotted line separating the occupied and unoccupied 
regions in Fig.~\ref{lce_1} is proportional to $\pm V_{\rm max}$; this 
clearly demonstrates the light cone like propagation.

The above scenario leads to a couple of important observations. First, 
when the system is observed stroboscopically one finds a linearly
spreading boundary separating the occupied and unoccupied regions. This 
emphasizes the existence of a $V_{\rm max}$ with which information (in this 
case, the bosons themselves) can propagate, thereby establishing an 
equivalent to the Lieb-Robinson limit \ct{lieb72} in a sinusoidally driven 
quantum system which is at a gapless quantum critical point 
in the absence of the driving term. Second, there can be situations when 
$V_{\rm max}$ vanishes in the asymptotic limit $n \to \infty$, which 
corresponds to a real space dynamical localization of the particles; this 
also implies that the system stops absorbing energy over a complete 
period even if it is being periodically driven. However, the 
particles would spread uniformly if the driving is stopped, which leads 
to the conclusion that this localization is indeed a result of 
the periodic driving.

Recently, there has been an experimental observation of 
light cone 
like spreading of ``two-point parity correlation" in an optical lattice
under a sudden quenching of the on-site interaction strength from a deep Mott 
insulating phase to the vicinity of a superfluid-Mott insulator boundary 
\ct{cheneau12}. The existence of an upper bound on the speed has been 
explained using the notion of the counter-propagation of quasiparticles 
(``holon" and ``doublon") generated due to the quench. Moreover, a dynamical 
localization-to-delocalization transition has been observed in a quantum 
kicked rotator, realized by placing cold atoms in a pulsed, far-detuned, 
standing wave, by measuring the number of zero velocity atoms under the 
influence of a quasiperiodic driving \ct{ringot00}. In connection to our 
work, the dynamical localization we predict can be experimentally observed by 
realizing the hard core boson model in an optical lattice with a sinusoidally 
varying alternating on-site potential and measuring the current 
stroboscopically starting from an initial current carrying ground state 
(obtained by applying a synthetic gauge potential \ct{lin09} to the 
one-dimensional optical lattice). A vanishing current in the large time limit 
($t=nT$ with $n\gg 1$) for a particular set of values of $V_0$ and $\om$ would 
signify the existence of a dynamically localized state. Similarly, the light 
cone propagation of a wave packet can also be realized by measuring the 
quasiparticle correlation function with time; the upper bound on the velocity, 
$V_{\rm max}$, can be determined from the first maximum of the correlation 
function as a function of time and distance. 

\vspace*{.6cm}
\section{Concluding remarks}
\label{concluding}

We have analyzed the behavior of a one-dimensional system of hard core bosons 
which have a nearest neighbor hopping amplitude $\ga = 1$ and are driven by 
a sinusoidally varying staggered potential with magnitude $V_0$ and frequency 
$\om$. We have derived the maximum group velocity $V_{\rm max}$ from the 
quasi-energies computed numerically from the Floquet operator. A number of 
analytical approximation methods have been used to study $V_{\rm max}$ in
different regions in the parameter space. Within the adiabatic approximation 
(which is valid when $\om \to 0$), we find that $V_{\rm max}$ is independent 
of $\om$ for small $V_0$ and scales as $1/V_0$ for large $V_0$. For large 
frequencies we use the Magnus expansion of the Floquet Hamiltonian and find 
that $V_{\rm max}=2$, independent of the magnitude of $V_0$. In this limit, 
the periodic perturbation vanishes on average and only the tight-binding 
part of the Hamiltonian contributes to the group velocity. In the limit 
$V_0 \gg \ga$, we show that the Floquet perturbation theory correctly predicts
the vanishing of $V_{\rm max}$ for $J_0(2V_0/\om)=0$; this dynamical 
localization is particularly prominent in the limit of large $V_0$ and $\om$ 
with $V_0/\om \sim 1$. In the other limit $V_0 \ll \ga$, there is a correction 
to the group velocity at first order in $V_0$ when the condition $4 \ga 
\cos k =m \om$ is satisfied. 

Finally, starting from an initial state where the hard core 
bosons are localized in one part of the chain, we demonstrate that the 
existence of $V_{\rm max}$ sets an upper bound on the speed with which 
particles propagate leading to a light cone like spreading of the particles 
in real space. The dynamically localized (or frozen) phase is characterized
by a vanishing $V_{\rm max}$.

None of the analytical methods work in the intermediate region when $V_0$ and 
$\om$ are both of the order of $\ga$ (shown by the central region in the
right panel of Fig.~\ref{lce}). An analysis of the behavior of $V_{\rm max}$ 
in this region may be an interesting subject for future research.

We conclude with the remark that the result presented here 
for a sinusoidal driving is not special to a one-dimensional model of hard 
core bosons (which is equivalent to a fermionic model in one dimension). 
A similar behavior of the stroboscopic group velocity, especially the 
dynamical localization, can be shown to occur in models with non-interacting 
fermions on a variety of higher dimensional hyper-cubic lattices.

\section*{Acknowledgments}
For financial support, D.S. thanks DST, India for Project No. 
SR/S2/JCB-44/2010, and A.D. acknowledges DST, India for Project No. 
SB/S2/CMP-19/2013. We acknowledge Arnab Das and Sthitadhi Roy for discussions.

\end{document}